\documentclass[aps,superscriptaddress,nofootinbib,eqsecnum,prd,notitlepage,twocolumn]{revtex4-1} 

\pdfoutput=1

\usepackage{amsfonts}
\usepackage{amsmath}
\usepackage{amssymb}
\usepackage{graphicx,color}
\usepackage{float}
\usepackage{hyperref}
\usepackage{subfigure}
\usepackage{dcolumn}
\usepackage{soul}
\usepackage{ulem}
\usepackage{verbatim}
\usepackage{bm}  
\usepackage{slashed}

\begin{document}

\title{Tilted Dirac cone effects and chiral symmetry breaking
  in a planar four-fermion model}

\author{Y. M. P. Gomes}
\email{yurimullergomes@gmail.com}
\affiliation{Departamento de F\'{\i}sica Te\'{o}rica, Universidade do
  Estado do Rio de
Janeiro, 20550-013 Rio de Janeiro, RJ, Brazil}

\author{Rudnei O. Ramos}
\email{rudnei@uerj.br}
\affiliation{Departamento de F\'{\i}sica Te\'{o}rica, Universidade do
  Estado do Rio de
Janeiro, 20550-013 Rio de Janeiro, RJ, Brazil}

\begin{abstract}

We analyze the chiral symmetry breaking in a planar four-fermion model
with non-null chemical potential, temperature and including the effect
of the tilt of the Dirac cone.  The system is modeled with a
$(2 + 1)$-dimensional Gross-Neveu-like interaction model in the context
of the generalized Weyl Hamiltonian and its phase structure is studied
in the mean-field and large-$N$ approximations.  Possible applications
of the results obtained, e.g., in connection to graphene, are
discussed. We also discuss the effect of an  external magnetic field
applied to the system, which can give rise to the appearance of the
anomalous Hall effect and that is expected to arise in connection with
two-dimensional Weyl and Dirac semimetals.  

\end{abstract}

\maketitle

\section{Introduction}

Condensed-matter systems, in particular, those with linearly dispersing
fermionic excitations, have been serving as perfect platforms for
testing predictions made by quantum field theories.  This interplay
between high-energy physics and condensed matter has been producing
new insights about the quantum phenomena.  This includes the
parallel between the Ginzburg–Landau model of superconductivity and
the spontaneous symmetry breaking in the Higgs sector of the standard
model of particle physics to the relation between topological
insulators and axion electrodynamics, with applications, e.g., in
two-dimensional
systems~\cite{Yu:2019opk,Gao:2017bqf,Rahmani:2015qpa,Grover:2013rc,Bernal:2019ook,Wu:2016oxw,Coleman:1974jh}.
The physics of graphene~\cite{Novoselov:2005kj} is of particular
relevance in this context, where the electrons can be treated in a
quasirelativistic way through the Dirac equation in
$(2 + 1)$ dimensions.

Although the relativistic character of the electrons in graphene is
intrinsically connected with the Lorentz symmetry, this is not the
case in most materials in condensed matter. There are condensed matter
systems where the dispersion in the vicinity of band touching points,
even though they can be generically linear and resemble the Weyl
equation, they lack Lorentz invariance.  Despite the fact that Lorentz
symmetry breaking in high-energy physics may emerge in specific
contexts~\cite{Kostelecky:1988zi,Mocioiu:2001fx,Mavromatos:2009mj,Mavromatos:2007xe,Colladay:1996iz,Colladay:1998fq,Kostelecky:2008bfz},
they face strong constraints in
general~\cite{Mattingly:2005re,Kostelecky:2015nma}.
Despite this, models exhibiting  Lorentz symmetry breaking have been
recently applied in the condensed-matter area, e.g., as a path to
model three-dimensional Weyl
semimetals~\cite{Grushin:2012mt,weyl3D2,weyl3D3,Wan:2011udc},
quasi-planar organic materials, and heterostructures made of a
combination of topological and normal
insulators~\cite{weyl1,weyl2,weyl3,weyl4}.       

These systems are described by Weyl-like Hamiltonians and,
therefore, these systems have quasiparticles that behave like Weyl
fermions~\cite{Weyl:1929fm}. These quasiparticles are by construction
massless and are more stable against gap formation in comparison to
Dirac ones. After several theoretical predictions, the first
experimental measurement related to this kind of
material~\cite{Xu:2015cga} shed light on the properties of the first
detected Weyl semimetal (for reviews, see, e.g.,
Refs.~\cite{Armitage:2017cjs,Yan:2017jgt}). 

In the present paper, we study how the gapless property associated with
Weyl fermions would hold against possible chiral symmetry breaking.
Here, the usual Weyl Hamiltonian used in the description of Weyl
semimetals~\cite{weyl1,weyl2} is extended by the introduction of a
four-fermion interaction. The properties of this system are then
studied under the effects of both a finite chemical potential and
temperature.

Let us recall that four-fermion interacting models, particularly
Gross-Neveu (GN)-type models~\cite{Gross:1974jv} in
$(2+1)$ dimensions, are of particular interest. This is because of
their simplicity and ability to capture the relevant physics of planar
fermionic systems in general and, thus, these types of models have been
extensively considered in the literature. These models can have either
a discrete chiral symmetry, $\psi \to \gamma_5 \psi$, or a continuous
one, $\psi \to \exp(i \alpha \gamma_5) \psi$. These types of models
have been employed to study, e.g., low-energy excitations of
high-temperature superconductors~\cite{Liu:1998mg}, while analogous
models with four-fermion interactions have also been used to study the
quantum properties of graphene~\cite{Drut:2007zx}.  Taken as an
effective low-energy description of the intrinsic physics of many
relevant condensed matter systems of interest, the four-fermion models
of the GN type have then served as motivation for many previous
studies (see, e.g.,
Refs.~\cite{Caldas:2008zz,Caldas:2009zz,Ramos:2013aia,Klimenko:2013gua,Ebert:2015hva,Ebert:2016ygm,Zhukovsky:2017hzo,Zerf:2017zqi,Fernandez:2021dfk}
for different examples of applications).  In particular, these models
can play a role in understanding the possibility of dynamical gap
generation in planar condensed matter systems, e.g.,  in
graphene~\cite{Drut:2007zx,Juricic:2009px,Herbut:2009vu,Rostami:2020set}.

The aim of this paper is then to study a Weyl-like Hamiltonian when it
is augmented by a four-fermion interaction of the GN type. We study
the interplay between the characteristic anisotropy and tilting of the
Dirac cone,  as typically considered in these models,  with both
temperature and chemical potential (i.e., the effects of doping) in
the phase structure of the resulting system.  This study is performed
in the context of the effective potential in the mean-field and
large-$N$ approximation. It is expected that this paper will be of
interest for learning some of the analogous effects that might be
important when considering fermionic quasiparticles and excitonic
bound states in planar fermionic systems of relevance.

The remainder of this paper is organized as follows. In
Sec.~\ref{sec2}, we show the relevant properties of two-dimensional
Dirac and Weyl semimetal systems. In Sec.~\ref{sec3}, we extend the
model by considering a four-fermion interaction and analyze the
effective potential for the averaged value for the chiral operator.
The effective potential is derived at the mean-field and large-$N$
level when taking into account the effects of both the anisotropy,
tilting of the Dirac cone, chemical potential and temperature. We
dedicate Sec.~\ref{sec4} to discuss our results and also some possible
applications. In Sec.~\ref{sec5}, we discuss some expected features
for this Weyl-type model when an external magnetic field is applied to
the system. {}Finally, in Sec.~\ref{conclusions}, we give some
additional discussions concerning our results and also our concluding
remarks.    Throughout this paper, we will be considering the natural
units where $\hbar = k_B = c=1$.

\section{Two-dimensional Weyl semimetals}
\label{sec2}

\begin{center}
\begin{figure}[!htb]
\includegraphics[width=7.5cm]{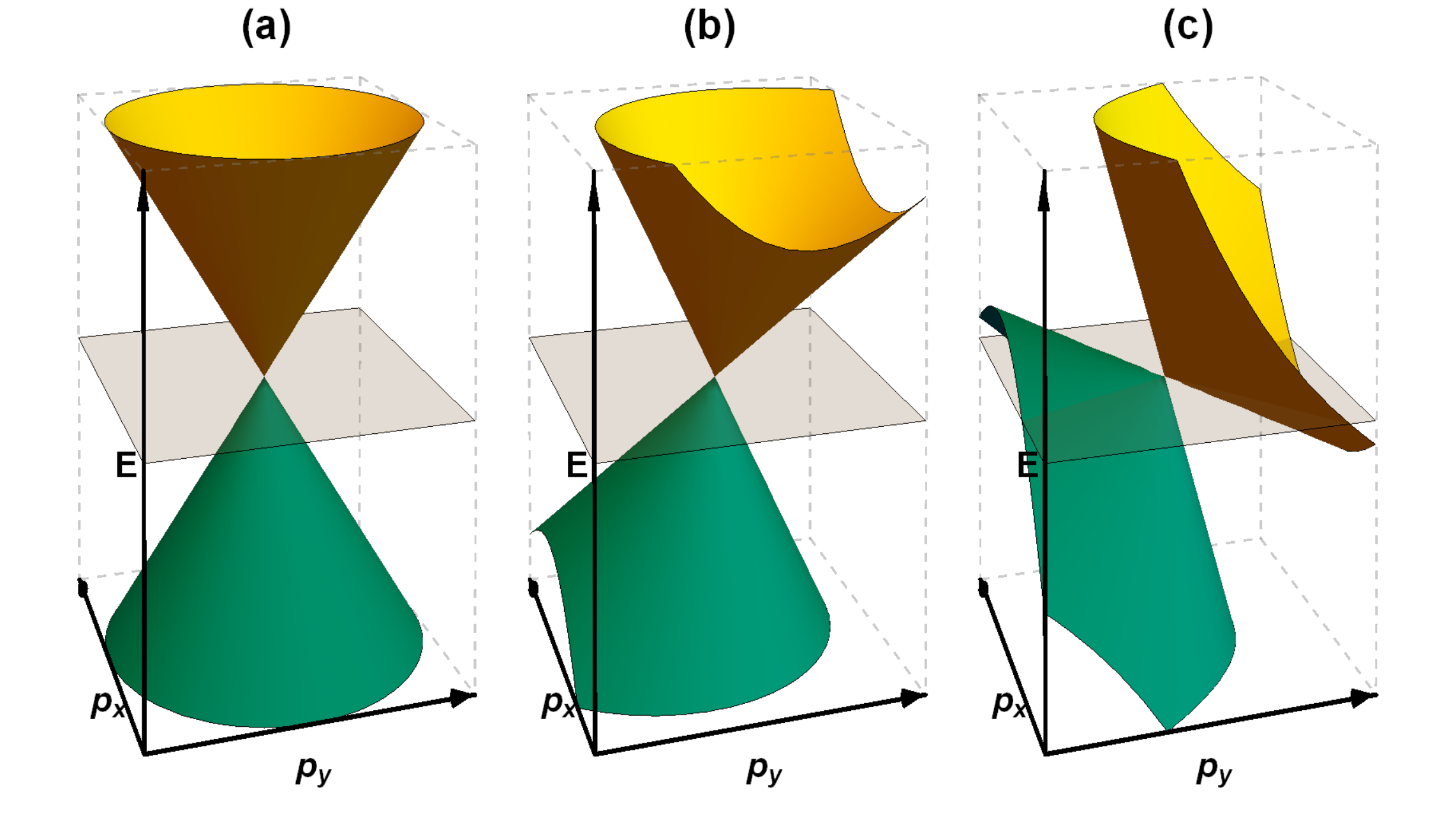}
  \caption{Illustration of a generic Dirac cone for an isotropic
    Dirac semimetal (a), type-I Weyl semimetal (b), and type-II Weyl
    semimetal (c).}
\label{fig1}
  \end{figure}
\end{center}

The generalized Hamiltonian which describes the electrons in the
two-dimensional Weyl semimetal (2D WSM) is given
by~\cite{weyl1,weyl2},
\begin{equation}\label{H1}
H_t({\bf p}) = v_F\left[ \left( {\bf t} \cdot {\bf p}\right) \tau^0 +
  \left(\xi_x p_x\right)\tau^x + \left( \xi_y p_y\right)\tau^y
  \right], 
\end{equation} 
where $v_F$ is the {}Fermi velocity, $\tau^0 = \mathbb{I}$ the
$2\times 2$ identity matrix, $\tau^x$ and $\tau^y$ are the Pauli
matrices, ${\bf t}$ is the vector defining the tilt of the Dirac cone
and ${\bm \xi} = (\xi_x,\xi_y)$ is the vector that describes the
anisotropic character of the crystalline structure.  The tilt vector
${\bf t}$ is related to the separation between the Dirac cones in the
Weyl semimetal. A consequence of the non-null tilt term in
Eq.~(\ref{H1}) is that the Dirac points, denoted by $D$ and $D'$, no
longer coincide with the Brillouin corners $K$ and $K'$ (see, e.g.,
Ref.~\cite{weyl2} for details and a review).  In particular, type-I
Weyl semimetals are characterized by $|{\bf t}|< 1$, while  type-II
ones by $|{\bf t}|> 1$ (see {}Fig.~\ref{fig1} for an
illustration). Although $|{\bf t}|$  is not restricted, to
maintain the physical meaning of the Hamiltonian Eq.~(\ref{H1}), it is
imposed that
\begin{equation}\label{cond1}
\sqrt{\left( \frac{t_x}{\xi_x}\right)^2 +\left(
  \frac{t_y}{\xi_y}\right)^2} = |\tilde{\bf t}| <1,
\end{equation}
where $|\tilde{\bf t}|$ is called the effective tilt parameter. When
the  condition given by Eq.~(\ref{cond1}) is not respected, the
isoenergetic lines are no longer ellipses, but instead they are
hyperboles~\cite{weyl1,weyl2}.  The above condition on the effective
tilt parameter will also become evident when we give our results for
the critical points associated with the chiral symmetry restoration  later
on. One should also notice that while $|\tilde{\bf t}|$ is restricted
to be smaller than unity, we can still have either the type-I or
type-II Weyl cases,   $|{\bf t}|< 1$ or $|{\bf t}|>  1$, respectively,
depending on the values for the the anisotropies $\xi_x$ and $\xi_y$.
The value of $|\tilde{\bf t}|$ depends on the material. {}For example,
in quinoid-type deformed graphene~\cite{weyl1}, it is of order of $0.6
\epsilon$, with the relative strain parameter $\epsilon <0.1$ for
moderate deformations. The degree of freedom described by the
Hamiltonian Eq.~(\ref{H1}) is that of a (massless) Weyl fermion.

The Hamiltonian spectrum obtained from Eq.~(\ref{H1}) is given by
\begin{equation}
E({\bf p}) = v_F \left[ {\bf t} \cdot {\bf p} + \lambda \sqrt{(\xi_x
    p_x)^2 +(\xi_y p_y)^2} \right],
\end{equation} 
where $\lambda =\pm 1$ define the conduction and valence bands,
respectively. Note that the condition  Eq.~(\ref{cond1}) ensures the 
association of $\lambda=+1$ to a positive and $\lambda=-1$ to a 
negative energy state~\cite{weyl1,weyl2}.
 In the untilted case $t_x=t_y = 0$ and isotropic one,
$\xi_x = \xi_y = 1$, we recover the spectrum of isotropic
graphene. The Hamiltonian Eq.~\eqref{H1} commutes with the chirality
operator given by
\begin{equation}
\mathcal{C} = \frac{(\xi_x p_x)\tau_x + (\xi_y
  p_y)\tau_y}{\sqrt{(\xi_x p_x)^2 +(\xi_y p_y)^2}},
\end{equation}
and the eigenvalues of $\mathcal{C}$ are given by $\alpha = \pm
1$. Due to the two-fold degeneracy of the Dirac cones $D$ and $D'$,
with representative index $\rho = \pm 1$, the band index can be
properly identified as $\lambda = \rho \alpha$.  Taking these
degeneracies into account, we can write a four-component Weyl spinor
$\psi$ and a Dirac-like Lagrangian density can be written as 
\begin{eqnarray}
\mathcal{L}_0&=& i  \bar{\psi} \left[ \left( \partial_t - v_F{\bf t}
  \cdot {\bm \partial} \right) \gamma^0 - v_F\sum_i(\xi_i
  \partial_i)\gamma^i  \right]  \psi   \nonumber\\ &=& i  \bar{\psi}
M^{\mu \nu} \gamma_\mu \partial_\nu \psi,
\label{lagr1}
\end{eqnarray}
where
\begin{equation}
\gamma^\mu = \tau^\mu \otimes \begin{pmatrix} 1 & 0 \\ 0 & - 1 
  \end{pmatrix},
\end{equation} 
$\tau^{\mu} = (\tau_3, i \tau_x, i\tau_y)$, and $\bar{\psi} =
\psi^\dagger \gamma^0$. The $\gamma$-matrices respect the algebra
$\{\gamma^\mu, \gamma^\nu\}= 2 \eta^{\mu \nu}$ (for details of this
representation for fermions in $(2+1)$-dimensions see, e.g.,
Ref.~\cite{Rosenstein:1990nm}). We also have that 
\begin{equation}  
M^{\mu \nu} = \begin{pmatrix} 1 & -v_F t_x & -v_F t_y\\ 0 & -v_F\xi_x
  & 0\\ 0 & 0 & -v_F\xi_y
\end{pmatrix}.
\end{equation}
One should note that $M^{\mu \nu}$ contains the parameters that
explicitly break the Lorentz symmetry. Note also that we do not
introduce the two-fold spin degeneracy here and, in case of interest,
we have to double the spinors $\psi \rightarrow \psi_s$ for $s=\pm 1$.
It is easy to show that the Lagrangian density Eq.~\eqref{lagr1} has a
discrete chiral symmetry given by $\psi \rightarrow \gamma_5 \psi$ and
$\bar{\psi} \rightarrow - \bar{\psi} \gamma_5 $
with~\cite{Rosenstein:1990nm}
\begin{equation}
i\gamma_5 = \begin{pmatrix} 0 & \mathbb{I} \\ -\mathbb{I} & 0
  \end{pmatrix}.
\end{equation} 

Though there are some different ways for adding a mass term in the
Lagrangian density Eq.~\eqref{lagr1} breaking the chiral
symmetry~\cite{Rosenstein:1990nm}, the simplest one is the mass term
$\bar{\psi} \psi$. In particular, this mass term can be generated
dynamically once the model given by Eq.~(\ref{lagr1}) is extended to
include a four-fermion interaction of the GN type. This extension is
considered next.
  
\section{Extending the model with a four-fermion interaction}
\label{sec3}

As already mentioned in the Introduction, the use of local
four-fermion interactions is well motivated in the literature. These
interactions can, for example, describe in an effective way the
interaction between electron and phonons in materials~\cite{phonons}.
{}Four-fermion-type interactions can also  describe the effects of
impurity/disorder~\cite{Zhao:2018jro,Wang:2018trs} and,  in the case
of the GN type of model, it has also been used in the honeycomb
lattice searching for the possibility of gap
opening~\cite{Drut:2009aj,Herbut:2006cs,Son:2007ja,quant1,Drut:2008rg,Juricic:2009px}.
In addition, these local interactions can be seen as an extension of the
BCS theory, effectively describing $s$-wave low-energy interactions, for
example.  {}For graphene, for example, in addition to the Coulomb
interaction, the effective continuum model  for quasiparticles is also
expected to contain contact four-fermion interaction terms, which
arise  from the original lattice tight-binding
model~\cite{tb1,Herbut:2006cs,tb2,tb3,Aleiner:2007va}.  All these
possible interactions, which might be important in opening a gap in 2D
systems, are in addition to the long-range Coulomb interaction.  There
have also been some works modeling the long-range Coulomb interaction
in an effective way (see, e.g., Ref.~\cite{Ebert:2015hva}), valid at
low energies, through a four-fermion  interaction. This could be the
case in superconducting materials where the mediating photon for the
electron-electron interaction acquires an effective mass. In this
case, for momentum less than the effective photon mass, the long-range
Coulomb interaction could be effectively approximated by a local
contact interaction.  In the present paper, we avoid attaching to the
interaction any of the above  possibilities and use the four-fermion
interaction solely as an effective interaction which can work in
producing a gap in the system. 

{}For the purpose of analyzing the metal-insulator phase transition in
the 2D WSM, the Lagrangian density Eq.~\eqref{lagr1} is extended to
include a four-fermion  interaction,
\begin{equation}\label{lagr2}
\mathcal{L} =  i \bar{\psi}  M^{\mu \nu}\gamma_\mu \partial_\nu \psi +
\frac{\lambda  v_F}{2N} (\bar{\psi} \psi)^2,
\end{equation}
where $\psi$ is a fermionic field with $N$ flavors (the sum over the
flavors implicit) and $\lambda$ is the coupling constant. Note that
the Lagrangian density Eq.~\eqref{lagr2} is still invariant under the
discrete chiral symmetry, unless $\langle \bar{\psi} \psi \rangle \neq
0$. Introducing an auxiliary field $\sigma$, the Lagrangian density
Eq.~\eqref{lagr2} can be equivalently rewritten as
\begin{equation}
\mathcal{L} =  \bar{\psi} ( i  M^{\mu \nu}\gamma_\mu \partial_\nu -
\sigma) \psi - \frac{N}{2  v_F \lambda} \sigma^2.
\label{lagr3}
\end{equation}
The Lagrangian densities Eqs. (\ref{lagr2}) and (\ref{lagr3}) are
completely equivalent. This becomes evident by noticing that the
Euler-Lagrange equation of motion for $\sigma$ and which is obtained
from Eq.~(\ref{lagr3}) is simply $\sigma = - v_F \lambda/N  \bar{\psi}
\psi$.  When this is substituted back in Eq.~(\ref{lagr3}), we recover
exactly Eq.~(\ref{lagr2}).  The GN Lagrangian density expressed in the
form of Eq.~(\ref{lagr3}) makes evident the chiral operator in terms
of the scalar field $\sigma$ and it is better suitable to study the
dynamical fermion mass generation in the
model~\cite{Rosenstein:1990nm}.

By integration of the fermionic degree of freedom in the mean-field
approximation, where $\sigma_c\equiv \langle \sigma \rangle$ is a
constant  background field, the effective potential for $\sigma_c$, at
one-loop order, is given by
\begin{equation}
V_{\rm eff} = \frac{N}{2 v_F \lambda }\sigma_c^2 + {\rm tr} \ln \left(
i  M^{\mu \nu}\gamma_\mu \partial_\nu - \sigma_c \right).
\label{Veff}
\end{equation}
Writing Eq.~(\ref{Veff}) in Euclidean momentum spacetime and taking
the trace, we find that
\begin{widetext}
\begin{eqnarray}
V_{\rm eff} = \frac{N}{2 v_F \lambda }\sigma_c^2 - 2N  T
\sum_{n=-\infty}^{+\infty} \int\frac{d^2 p}{(2 \pi )^2}\ln \left[
  (\omega_n + i v_F {\bf t} \cdot {\bf p} +i\mu)^2 +v_F^2  ({\bm
    \xi}\cdot {\bf p})^2+ \sigma_c^2 \right].
\label{Veff2}
\end{eqnarray}
\end{widetext}
where $\omega_n = (2 n +1) \pi /\beta$ (with $n\in \mathbb{Z}$, $\beta
= 1/T$ and $T$ is the temperature of the system) are the Matsubara's
frequencies for fermions and $\mu$ is the chemical potential. Note
that the chemical potential can be interpreted as to account for the
extra density of electrons that is supplied to the system by the
dopants and, hence, is directly related to the doping
concentration. It is also worth pointing out that Eq.~(\ref{Veff2}) is
an exact result in the large-$N$
approximation~\cite{Rosenstein:1990nm}. Even though for practical
purposes $N$ is finite (e.g., in graphene $N=2$), we will assume that
Eq.~(\ref{Veff2}) still provides a sufficiently good approximation as
generally assumed in these four-fermion type models~\cite{Caldas:2009zz,Ramos:2013aia,Klimenko:2013gua,Ebert:2015hva}.
We will comment more on the validity of the large-$N$ approximation
used here when we discuss our results in the next section.

After summing over the Matsubara's frequencies in Eq.~(\ref{Veff2}),
we find that the effective potential is given by
\begin{widetext}
\begin{equation}\label{VeffTmu}
V_{\rm eff}(\sigma_c, \mu,T) = \frac{N}{2 v_F \lambda} \sigma_c^2 -2N
\int \frac{d^2 p}{(2 \pi )^2} \left\{E_\sigma + \frac{1}{\beta}
\ln\left[ 1 + e^{-\beta( E_{\sigma} + \mu)} \right] +  \frac{1}{\beta}
\ln \left[1 + e^{-\beta (E_{\sigma} - \mu)}\right] \right\},
\end{equation}
\end{widetext}
where 
\begin{equation}
E_{\sigma} =   v_F{\bf t} \cdot {\bf p} +\sqrt{ v_F^2\tilde{\bf p}^2 +
  \sigma_c^2},
\end{equation}
with $\tilde{\bf p}^2 = \tilde{p}_x^2 +\tilde{p}_y^2\equiv ({\bm
  \xi}\cdot {\bf p})^2$.

The so-called gap equation is given by 
\begin{equation}
\frac{\partial V_{\rm
    eff}}{\partial\sigma_c}\Bigr|_{\sigma_c={\bar\sigma}_c} = 0,
\label{gap}
\end{equation} 
which gives
\begin{eqnarray}
\!\!\!\!\!\! 1 &=&   2 \lambda v_F \int \frac{d^2 p}{(2 \pi )^2}
\frac{1}{\sqrt{ v_F^2\tilde{\bf p}^2 + {\bar\sigma}_c^2}} \nonumber
\\ &\times & \left[  1- \frac{1}{e^{\beta( E_{\sigma} + \mu)} +1}  -
  \frac{1}{e^{\beta( E_{\sigma} - \mu)} +1}
  \right]\Bigr|_{\sigma_c={\bar\sigma}_c},
\label{gapTmu}
\end{eqnarray}
along with the trivial solution ${\bar\sigma}_c=0$.

Special cases of the above equations will be analyzed next.

\subsection{The $T=\mu=0$ case}

Note that the effective potential Eq.~(\ref{VeffTmu}) is divergent.
Considering the limit $\beta \rightarrow \infty$ and $\mu \rightarrow
0$ in Eq.~(\ref{VeffTmu}) and performing the (divergent) momentum
integral with a cutoff $\Lambda$, we can define the renormalization
condition for the four-fermion interaction $\lambda_R$
as~\cite{Ramos:2013aia,Vshivtsev:1996ri}
\begin{eqnarray}
\frac{1}{\lambda_R(m)} &=& \frac{ v_F}{N} \frac{d^2 V_{\rm
    eff}(\sigma_c)}{d \sigma_c^2}\Bigr|_{\sigma_c=m} \\ \nonumber & =
&\frac{1}{\lambda} + \frac{2m}{ \pi \xi_x \xi_y v_F} - \frac{\Lambda}{
  \pi \xi_x \xi_y v_F},
\end{eqnarray}
where $m$ is a regularization scale. Next, defining the
renormalization invariant coupling $\lambda_R$ as
\begin{equation}
\frac{1}{\lambda_R}= \frac{1}{\lambda_R(m)}-\frac{2m}{ \pi \xi_x \xi_y
  v_F},
\end{equation}
we arrive at the renormalized effective potential as given by
\begin{eqnarray}
V_{\rm eff,R}(\sigma_c) &=&\frac{N}{2   v_F \lambda_R} \sigma_c^2
\nonumber \\ & + &\frac{N}{ 3 \pi\xi_x  \xi_y v_F^2} |\sigma_c|^3.
\label{VeffR0}
\end{eqnarray}
When $\lambda_R <0$, the nontrivial vacuum solution of
Eq.~(\ref{VeffR0}) is given by 
\begin{equation}
{\bar\sigma}_c = \sigma_0\equiv  \frac{ v_F \pi \xi_x \xi_y}{
  |\lambda_R| },
\label{sigma0}
\end{equation} 
while for $\lambda_R>0$ we have that ${\bar\sigma}_c=0$.
Note that
\begin{equation}
V_{\rm eff,R}(\sigma_c=\sigma_0) =\frac{ N (\pi \xi_x \xi_y)^2 v_F}{6
  \lambda_R^3}, 
\end{equation}
and, hence, Eq.~(\ref{sigma0}) corresponds to the true minimum of the
system.  We can also see that the value of $\sigma_0$ is modified by
the dependence on the anisotropy constants, $\xi_x,\, \xi_y$, and we
recover the usual result~\cite{Caldas:2009zz,Ramos:2013aia} in the
isotropic limit: $\xi_x,\,\xi_y \rightarrow 1$. One can further notice
here, in the zero temperature  and zero chemical potential limits,
that the tilt parameter ${\bf t}$ does not contribute to the effective
potential.

\subsection{The $T=0$ and $\mu \neq 0$ case}

At zero temperature but with a non-null chemical potential, from the
effective potential Eq.~(\ref{VeffTmu}) and considering
Eq.~\eqref{cond1}, one obtains that
\begin{eqnarray}
V_{\rm eff,R}(\sigma_c,\mu) &=& V_{\rm eff,R}(\sigma_c)  \nonumber
\\ &-& 2 N \int \frac{d^2 p}{(2 \pi )^2}\left( \mu - E_\sigma \right)
\Theta(\mu-E_\sigma), \nonumber \\
\label{VeffT0mu}
\end{eqnarray}
where $\Theta(x)$ is the standard Heaviside function. Even  though we
cannot explicitly make the integration in the momentum in
Eq.~(\ref{VeffT0mu}) analytically, we can  analyze it when the
effective tilt parameter $|\tilde{\bf t}|$ is small.  This is useful
to obtain an understanding of the effect of  $|\tilde{\bf t}|$.  {}For
a small effective tilt parameter, $|\tilde{\bf t}|\ll 1$, we can find
for Eq.~(\ref{VeffT0mu}) the approximated result
\begin{eqnarray}
V_{\rm eff,R}(\sigma_c,\mu) &=& V_{\rm eff,R}(\sigma_c) -  \frac{N}{2
  \pi\xi_x \xi_yv_F^2}\Theta(\mu^2 - \sigma_c^2) \nonumber \\ &\times
&\left[\frac{1}{3} \left(\mu^3 -3 \sigma_c^2 \mu +2
  |\sigma_c|^3\right)  + \frac{1}{2} |\tilde{\bf t}|^2 \mu(\mu^2 -
  \sigma_c^2) \right] \nonumber \\ &+&  O(|\tilde{\bf t}|^4).
\nonumber \\
\label{VeffT0mu2}
\end{eqnarray}

\begin{center}
\begin{figure}[!htb]
\subfigure[]{\includegraphics[width=7.5cm]{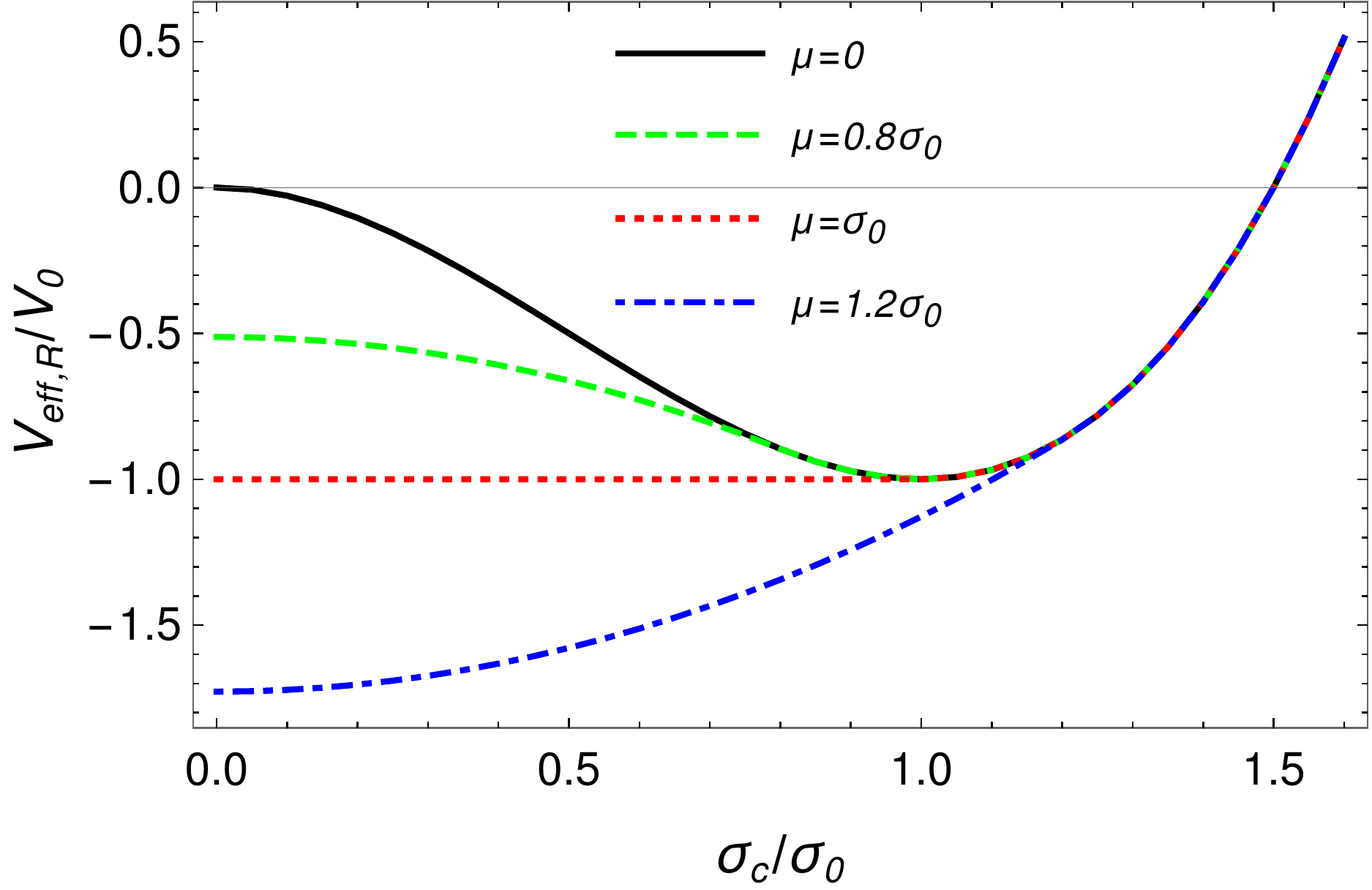}}
\subfigure[]{\includegraphics[width=7.5cm]{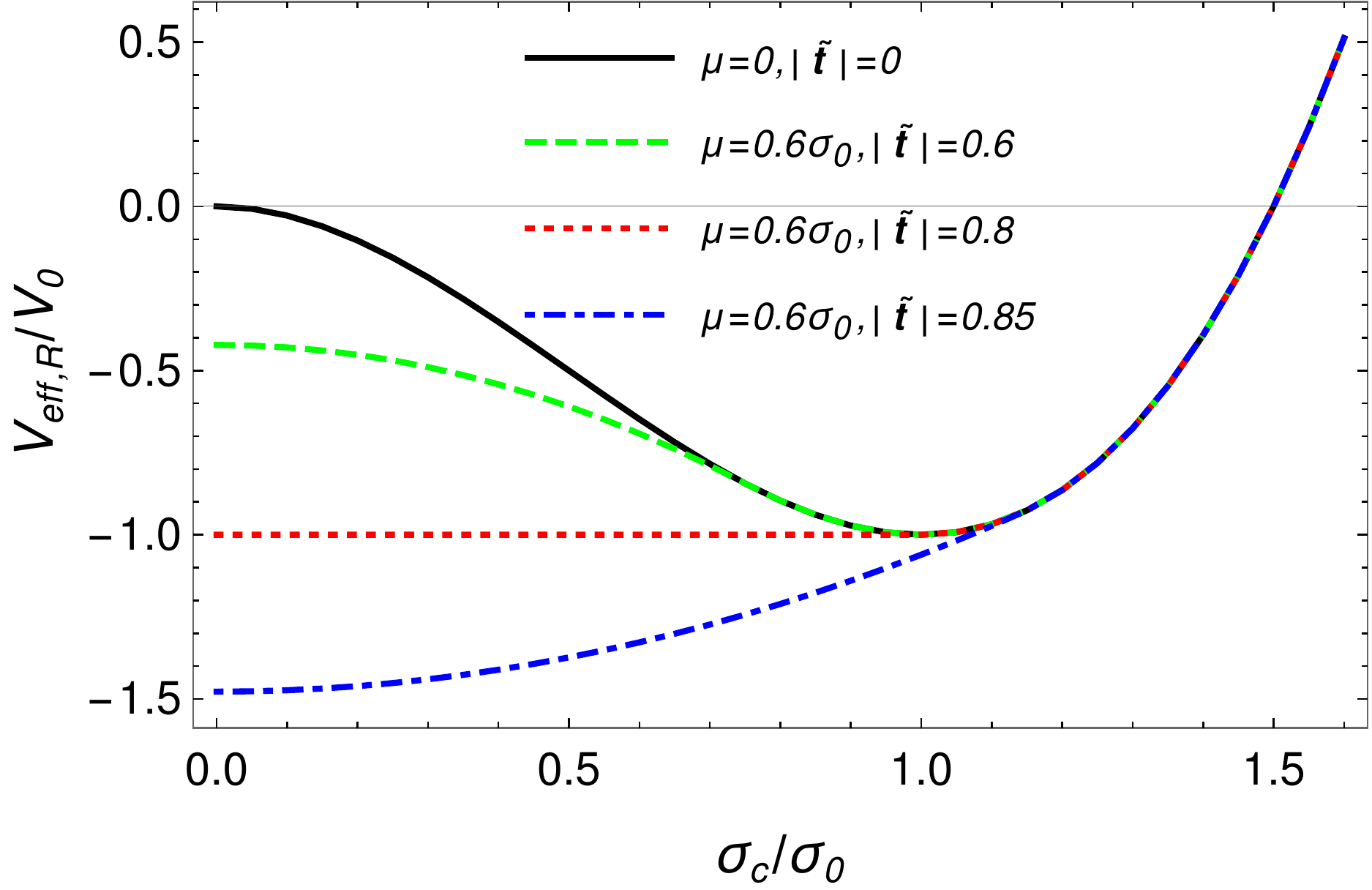}}
\caption{Normalized plot of the renormalized effective potential at
  zero temperature, Eq.~(\ref{VeffT0mu}), in units of $V_0\equiv N
  \sigma_0^3/(6 \pi v_F^2 \xi_x\xi_y)$. Panel (a) shows the effective
  potential when $|\tilde{\bf t}|=0$ and for different values for the
  chemical potential. Panel (b) shows the effect a nonvanishing
  $|\tilde{\bf t}|$ in the shape of the effective potential for the
  chiral order parameter.}
\label{fig2}
\end{figure}
\end{center}

Equation ~(\ref{VeffT0mu2}) shows that the effect of the effective tilt
parameter is to enhance the chemical potential. This is confirmed by
the results shown in {}Fig.~\ref{fig2}. {}Figure~\ref{fig2} shows the
effective potential as a function of $\sigma_c$ obtained from the
numerical integration of Eq.~(\ref{VeffT0mu}) and for some
representative values of $\mu$ and $|\tilde{\bf t}|$.   By comparing
the {}Figs.~\ref{fig2}(a) and ref{fig2}(b), we can see that the
effect of the  effective tilt parameter is to lower the point of
chiral  symmetry restoration. In the presence of the tilt the chiral
symmetry is restored at a lower value for the chemical potential,
which is in accordance to what  we find from the approximated
expression Eq.~(\ref{VeffT0mu2}).  This can be confirmed by
determining the critical point $\mu_c$ for which the chiral symmetry
gets restored.

To derive the critical point $\mu_c$, let us first recall that at a
vanishing effective tilt parameter, $|\tilde{\bf t}| =0$, the chiral
symmetry is restored through a first-order phase transition as has
been shown in many previous references (see, e.g.,
Refs.~\cite{Klimenko:1987gi,Rosenstein:1988dj,Rosenstein:1988pt}).  We
can see this explicitly happening in {}Fig.~\ref{fig2} (a).  In
the absence of the tilt parameter, at $\mu=0$ the effective potential
displays a maximum at $\sigma_c=0$ and a minimum at
$\sigma_c=\sigma_0$.  As the chemical potential increases, the minimum
remains located at $\sigma_c=\sigma_0$ and with an unchanged value for
the effective potential.  However, the value of the effective
potential at the maximum decreases.  This continues to happen until
the value $\mu=\sigma_0$ is considered. When $\mu=\sigma_0$, we have
that  $V_{\rm eff,R}(\sigma_c=\sigma_0,\mu=\sigma_0) = V_{\rm
  eff,R}(\sigma_c=0,\mu=\sigma_0)$ and the minimum and maximum of the
effective potential become degenerate.  In fact, at $\mu=\sigma_0$ the
whole range $0\leq \sigma_c \leq \sigma_0$ becomes degenerate and are
minimum points of the effective potential. As the chemical potential
increases further, $\mu> \sigma_0$, there will be only one minimum for
the effective potential and which is located at $\sigma_c=0$. The
chiral symmetry then becomes restored for $\mu>\sigma_0$. The chiral
order parameter, which is represented by the minimum of the effective
potential, will then jump discontinuously from $\sigma_c=\sigma_0$ to
$\sigma_c=0$ as we change the chemical potential from $\mu<\sigma_0$
to $\mu>\sigma_0$, with $\mu=\sigma_0$ representing the critical value
for the chemical potential. The discontinuous behavior for the order
parameter characterizes the transition as a first-order one.  In the
presence of the effective tilt parameter, i.e.,  $|\tilde{\bf t}| \neq
0$, we can see from {}Fig.~\ref{fig2}(b) that the same trend
remains.  However, in this case we have that the first-order
transition happens at a  lower value for the chemical potential. In
this sense, we can say that the presence of the nonvanishing tilt
parameter facilitates the chiral symmetry restoration. 

According to the above discussion, the critical point can then be
determined by the  condition $V_{\rm eff,R}(\sigma_c=0,\mu_c)=V_{\rm
  eff,R}(\sigma_c=\sigma_0,\mu_c)$.  Using Eq.~(\ref{VeffT0mu}), we
then find that the critical chemical potential as a function of the
effective tilt parameter is given by
\begin{equation}
\mu_c = \left(1-|\tilde{\bf t}|^2\right)^{1/2} \sigma_0.
\label{muc}
\end{equation}

\begin{center}
\begin{figure}[!htb]
\includegraphics[width=7.5cm]{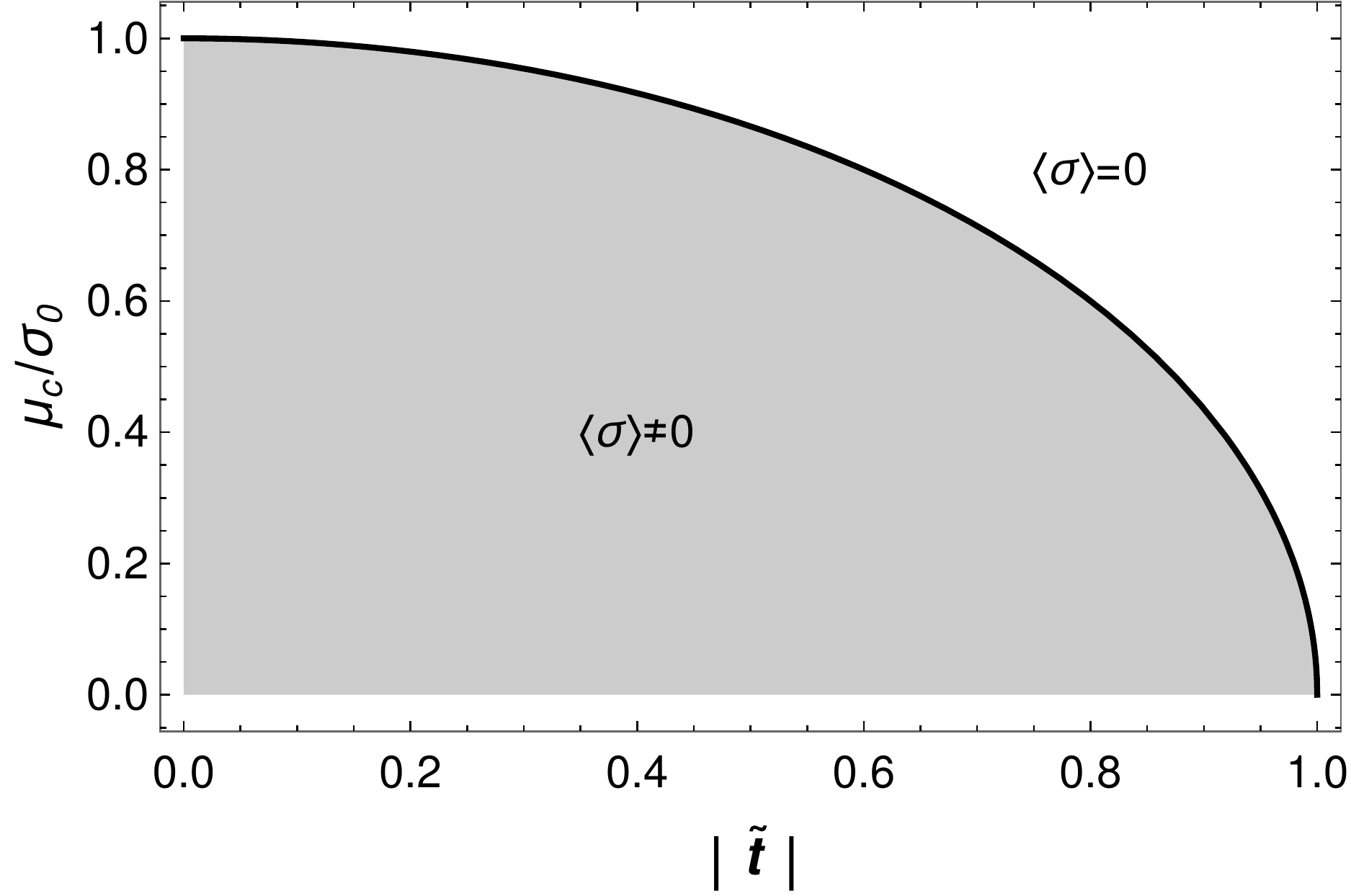}
\caption{The critical chemical potential $\mu_c$ (in units of
  $\sigma_0$) as a function of $|\tilde{\bf t}|$. }
\label{fig3}
\end{figure} 
\end{center}  

At the critical value $\mu_c$, the potential losses its non trivial
minimum, i.e., ${\bar\sigma}_c(\mu=\mu_c)=0$.  The critical value
$\mu_c$  as a function of the effective tilt parameter is shown in
{}Fig.~\ref{fig3}.    {}For $|\tilde{\bf t}|=0$, we recover the known
result $\mu_c = \sigma_0$ for the GN model in (2+1) dimensions.  The
result for the critical $\mu_c$  shown in Eq.~(\ref{muc}) clearly
shows that in the presence of a non-null tilt parameter, the gap
${\bar\sigma}_c$ will vanish at a smaller doping and, hence,
facilitating the chiral symmetry restoration. 

\begin{center}
\begin{figure}[!htb]
\includegraphics[width=7.5cm]{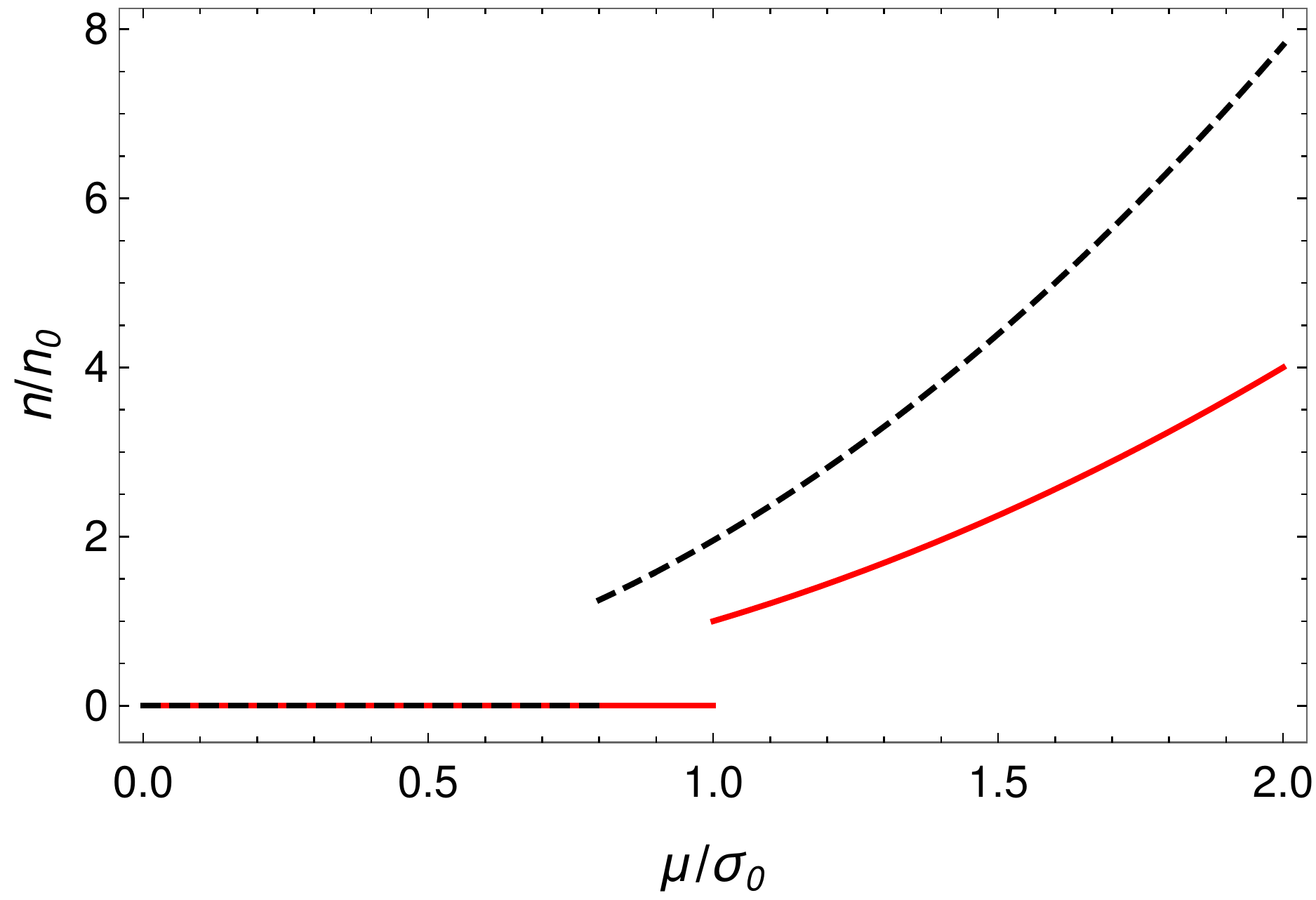}
	\caption{Density at zero temperature [in units of $n_0\equiv N
          \sigma_0^2/(2 \pi v_F^2 \xi_x\xi_y)$] as a function of the
          chemical potential for   $|\tilde{\bf t}|=0$ (solid line)
          and for  $|\tilde{\bf t}|=0.6$ (dashed line). }
\label{fig4}	
	\end{figure}
\end{center}

Note that the tilt parameter also affects the {}Fermi momentum, which
is now given by
\begin{equation}
\tilde{p}_F = \frac{\sqrt{\mu^2-\left[1-|\tilde{\bf
        t}|^2\cos^2(\theta)\right]\sigma_c^2}-\mu |\tilde{\bf
    t}|\cos(\theta)} {v_F\left[1-|\tilde{\bf
      t}|^2\cos^2(\theta)\right]},
\label{pF}
\end{equation}
where $\theta$ is the angle between ${\bf p}_F$ and the tilt vector
${\bf t}$.  We see that for $-\pi/2 < \theta < \pi/2$, the tilt
parameter acts towards increasing  the {}Fermi surface. This is also
reflected on the behavior of the charge density $n$, which is defined
as
\begin{equation}
n = - \frac{\partial V_{\rm eff,R}}{\partial
  \mu}\Bigr|_{\sigma_c=\bar{\sigma}_c},
\end{equation}
and, using Eq.~\eqref{VeffT0mu}, it gives the result 
\begin{eqnarray}
\label{density}
n(\sigma_c,\mu) &=& N\frac{\mu^2 - \left(1-|\tilde{\bf
    t}|^2\right)\bar{\sigma}_c^2}{2 \pi \xi_x \xi_y v_F^2
  \left(1-|\tilde{\bf t}|^2\right)^{3/2}} \Theta[\mu^2 -
  \left(1-|\tilde{\bf t}|^2\right)\bar{\sigma}_c^2].  \nonumber \\
\end{eqnarray} 
At the critical value $\mu_c$ given by Eq.~(\ref{muc}), we have that
$n(\bar{\sigma}_c=0,\mu_c)=N\sigma_0^2/(2 \pi \xi_x \xi_y v_F^2
\sqrt{1-|\tilde{\bf t}|^2})$, which is always larger than in the
absence of the tilt parameter. This is explicitly seen also in
{}Fig.~\ref{fig4}, where Eq.~(\ref{density}) is shown as a function of
the chemical potential, for a non-tilted Dirac cone case and also for
a tilted one. The discontinuity in the density here is a consequence
of the first-order phase transition happening at the critical point
$\mu_c$, where the chiral order parameter jumps discontinuously from
$\bar{\sigma}_c= \sigma_0$ to $\bar{\sigma}_c=0$ when going from $\mu<
\mu_c$ to $\mu>\mu_c$. 

\subsection{$T \neq 0, \mu = 0$}

Let us now analyze the case of the chiral symmetry restoration at
finite temperature, but in the absence of the chemical
potential. Taking $\mu=0$ in Eq.~\eqref{VeffTmu} and using
Eq.~(\ref{VeffR0}), we then obtain that the renormalized effective
potential at finite temperature is given by
\begin{eqnarray}
\!\!\!\!\!\!\!\!\!\!\!V_{\rm eff}(\sigma_c, \mu=0,T) &=& \frac{N}{2
  v_F \lambda_R} \sigma_c^2 + \frac{N}{ 3 \pi\xi_x  \xi_y v_F^2}
|\sigma_c|^3 \nonumber \\ &-&\frac{4N}{\beta} \int \frac{d^2 p}{(2 \pi
  )^2} \ln\left( 1 + e^{-\beta E_{\sigma}} \right).
\label{VeffTmu0}
\end{eqnarray}
The situation in this case can again be compared to the transition
pattern in the GN model in (2+1) dimensions when the tilt effect is
absent~\cite{Klimenko:1987gi,Rosenstein:1988dj,Rosenstein:1988pt}.  In
that case, the chiral order parameter $\bar{\sigma}_c$ is shown to
change continuously from the value $\sigma_0$ at $T=0$ to
$\bar{\sigma}_c=0$ at a critical temperature $T_c$, which then
characterizes the phase transition to be second order.  The presence
of the tilt does not change this transition pattern, but as in the
previous case of $T=0$ and $\mu\neq 0$, it will lower the critical
$T_c$ with respect to the case when $|\tilde{t}|=0$ as we now show.  

\begin{center}
\begin{figure}[!htb]
\includegraphics[width=7.5cm]{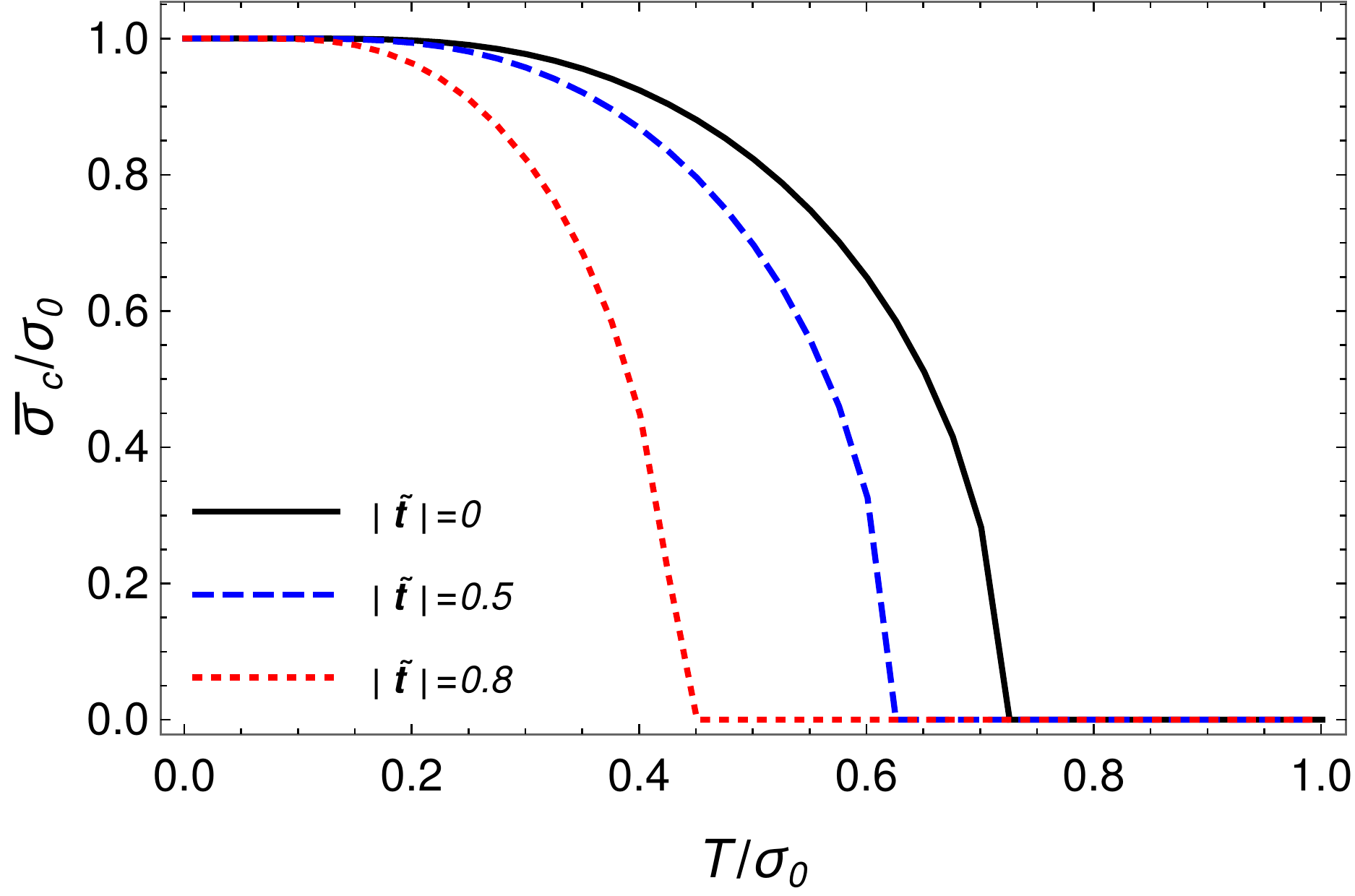}
\caption{The chiral order parameter  $\bar{\sigma}_c$ as a function of
  the temperature and for different values for the effective tilt
  parameter $|\tilde{\bf t}|$ when $\mu = 0$. }
\label{fig5}
\end{figure}
\end{center}  

\begin{center}
\begin{figure}[!htb]
\includegraphics[width=7.5cm]{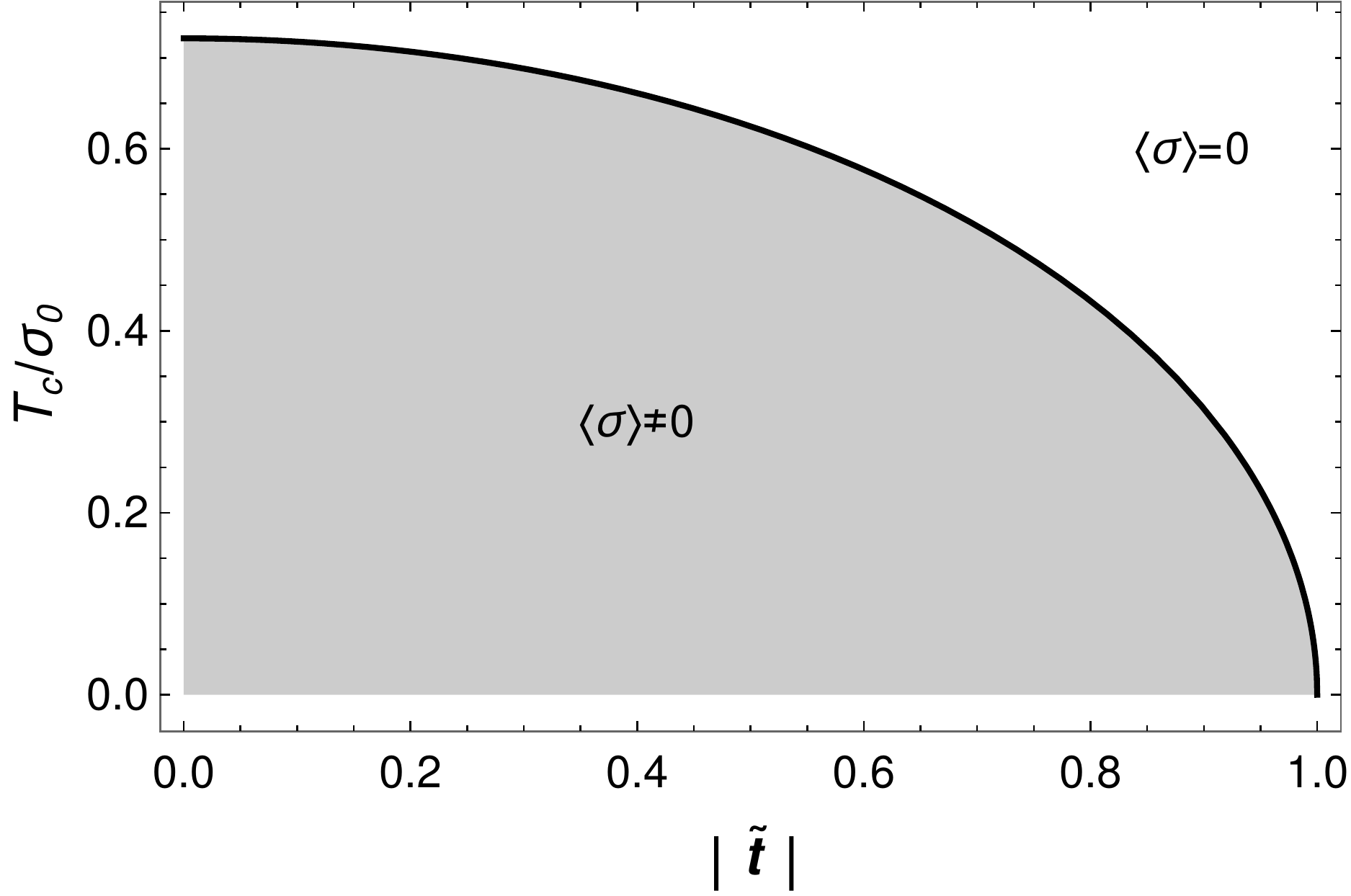}
\caption{The critical temperature as a function of $|\tilde{\bf t}|$
  for $\mu = 0$. }
\label{fig6}
\end{figure}
\end{center}

The chiral order parameter is obtained from the solution of the saddle point equation
\begin{equation}
\frac{\partial V_{\rm
    eff,R}}{\partial\sigma_c}\Bigr|_{\sigma_c={\bar\sigma}_c} = 0,
\label{gapR}
\end{equation} 
obtained from the renormalized effective potential. The result is
explicitly shown in {}Fig.~\ref{fig5}. In  {}Fig.~\ref{fig5}, it is
shown the chiral order parameter  $\bar{\sigma}_c$ as a function of
the temperature and for different values for the effective tilt
parameter. We note that $\bar{\sigma}_c$ changes continuously with the
temperature. The larger is the effective tilt parameter, the smaller
is $T_c$.  The dependence of the critical temperature with the
effective tilt parameter is analytical and can be derived directly
from Eq.~(\ref{gapR}) when setting $\bar{\sigma}_c=0$ in that
equation.  The solution obtained for $T_c$ is then found to be given
by
\begin{equation}
T_c= \sqrt{1-|\tilde{\bf t}|^2} \frac{\sigma_0 }{2 \ln 2}.
\label{Tc}
\end{equation}
The behavior of $T_c$ at $\mu=0$ and as a function of the tilt
parameter is shown in {}Fig.~\ref{fig6}.

When $|\tilde{\bf t}|=0$, we get the known result for the GN model in
(2+1)-dimensions~\cite{Klimenko:1987gi,Rosenstein:1988dj,Rosenstein:1988pt},
where $T_c= \sigma_0/(2 \ln 2)$.  Just like for the previous result
for the critical chemical potential, Eq.~(\ref{muc}), we also see here that the larger the effective tilt parameter, the lower the
critical temperature. Once again we see that the effect of the tilt is
to facilitate the chiral symmetry restoration, i.e., it can be reached
at a lower temperature than in the absence of the tilt. 

\subsection{$T \neq 0, \mu \neq 0$}

{}Finally, let us consider the case where both temperature and
chemical potential effects are included. The effective potential in
this case is given by Eq.~\eqref{VeffTmu}.  The critical curve for
which ${\bar\sigma}_c(T_c,\mu_c)=0$ is obtained again from the saddle
point equation derived from the renormalized effective potential. In
this case,  by setting  $\bar{\sigma}_c=0$ in the gap equation ~(\ref{gapTmu}), we explicitly find
\begin{equation}
\sigma_0 + \frac{\mu_c-2T_c \ln\left(e^{\mu_c/T_c}+1
  \right)}{\left(1-|\tilde{\bf t}|^2\right)^{1/2}}=0,
\label{Tcmuc}
\end{equation}
which reproduces the result Eq.~(\ref{muc}) in the zero temperature
limit, while at zero chemical potential it leads to the critical
temperature as given by Eq.~(\ref{Tc}).

\begin{center}
\begin{figure}[!htb]
\includegraphics[width=7.5cm]{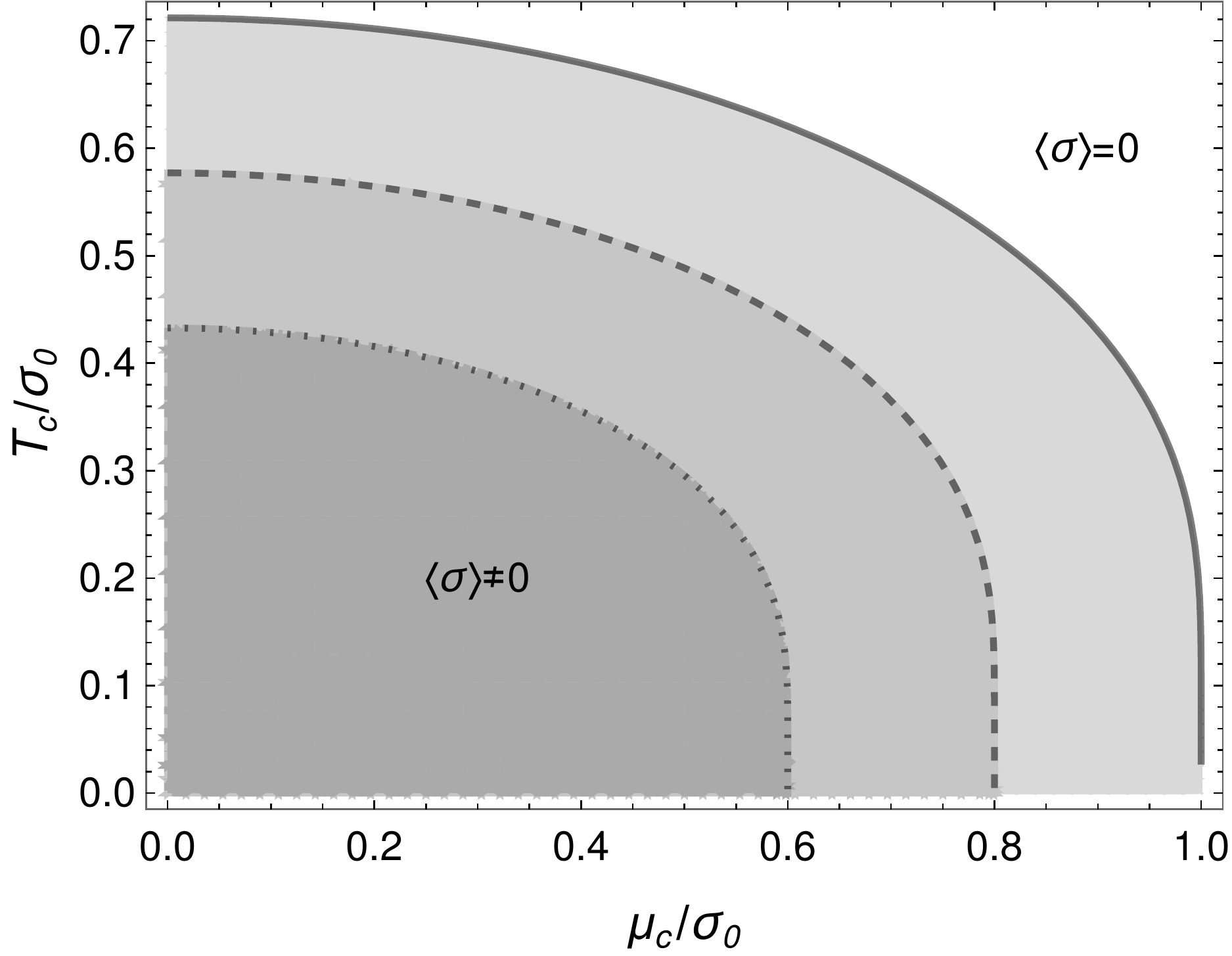}
\caption{Critical curve ${\bar\sigma}_c(\mu_c, T_c) = 0$ for
  $|\tilde{\bf t}| = 0$ (solid line) and for $|\tilde{\bf t}|= 0.6$
  (dashed line) and $|\tilde{\bf t}|=0.8$ (dotted line). }
\label{fig7}
\end{figure}
\end{center}  

The overall behavior of the critical curve obtained from
Eq.~(\ref{Tcmuc}) is shown in  {}Fig.~\ref{fig7}. One sees from this
figure that the larger the effective tilt parameter, the smaller
the region for chiral symmetry breaking. The critical curve in the
$(T,\mu)$ plane corresponds to a second-order chiral phase transition
line. As $T\to 0$, we reach the critical point $\mu_c$ as given by
Eq.~(\ref{muc}), which as already discussed above, corresponds to a
first-order chiral transition point.

\section{Some discussions of the results and applications}
\label{sec4}

As shown from the previous section, in particular when considering
Eqs.~(\ref{muc}) and (\ref{Tc}), the results depend directly on the
chiral order parameter $\sigma_0$. At the same time, $\sigma_0$, which
introduces a mass gap in the system, is affected by the anisotropy
through  the geometric mean {}Fermi velocity, $v_F^* = \sqrt{\xi_x
  \xi_y} v_F$.  The difference between the effective {}Fermi velocity
$v_F^*$ and the usual one in fully isotropic materials (considering
the case of, e.g., graphene, where $v_F \approx c/300$) can alter
significantly the critical temperature and critical chemical
potential. This is because the dependence of $v_F^*$ on the
anisotropies does not have a strong physical bound such as the one
observed by the effective tilt parameter $|\tilde{\bf t}|$ and given
by Eq.~\eqref{cond1}. 

It is worth providing some estimates, even though they might be
rough ones, based on results obtained from some current planar systems
of interest, particularly organic conductors. {}For illustration
purposes,  let us consider the case of the recent
study~\cite{hirata2016,Hirata:2021qdi} for the two-dimensional organic
conductor $\alpha - (\text{BEDT-TTF})_2 I_3$. {}From the data provided
in Refs.~\cite{hirata2016,Hirata:2021qdi}, the tilt vector ${\bf t}$
and anisotropy ${\bm \xi}$ vector can be estimated to be given (when
converted to our notation), respectively by  ${\bf t} \simeq (-1.7,
0.3) \times 10^{-4} $ and ${\bm \xi} \simeq (2.3, 2.4) \times
10^{-4}$, where we have assumed a graphene like {}Fermi velocity as a
base one (i.e., in the isotropic case). Thus, the effective {}Fermi
velocity is $v_F^* \approx 0.02 v_F$ and the effective  tilt parameter
can be estimated as $|{\bf \tilde{t}}| \approx 0.76$ . Therefore, from
Eqs.~(\ref{muc}) and (\ref{Tc}), the critical chemical potential and
critical temperature both get reduced by a factor $\sqrt{1-|\tilde{\bf
    t}|^2} \xi_x \xi_y \approx 3.6 \times 10^{-8}$ with respect to the
untilted and isotropic critical values. This is quite a substantial
reduction.  This is to be contrasted with the case of quinoid-type
graphene under uniaxial strain~\cite{weyl1}, where $v_F^* \approx v_F$
and the critical temperature and chemical potential of this system can
be estimated to decrease by a factor $(1-0.02 \epsilon^2)$, where
$\epsilon$ is the relative strain. Given that $\epsilon < 1$ for
moderate strain, this leads to a much smaller suppression of $\mu_c$
and $T_c$.   Thus, our results indicate that planar organic conductors
can remain gapless down to very small temperatures and dopings as a
result of the combined effects of anisotropy and the Dirac cone tilt.

There can also be other possible effects of the tilt and anisotropy of
the model.  Since both the tilt and the anisotropies break the
rotational symmetry,  there is a possibility that the gap generated is 
anisotropic in momentum space.  The study of such anisotropic
solutions and how they would contribute to the problem we have studied
would require, however, a study going beyond the proposed methods that
we have used in the present paper. In our paper, we have focused on the
effects of the tilt and anisotropy in the chiral symmetry breaking and
restoration.  The use of the effective potential, by definition, only
considers field configurations (represented here by the chiral
symmetry breaking order parameter $\sigma_c$) that are space and time
independent. There can be, nevertheless, other contributions to the
functional partition function that could contribute and that are
solutions of the (nonhomogeneous) field equations. We expect, at least
for the present model, those sorts of solutions would  cost more energy
though (e.g., due to the gradient energy terms) when compared to the
constant (homogeneous) field solution contributing to the partition
function (and, hence, to the effective potential). It would, of
course, be important to explore such solutions, perhaps  looking for
perturbations to the homogeneous background field solution we have
considered. This, of course, would also require a calculation
departing from the mean-field  approximation we have considered. 

{}Finally, let us comment on the reliability of the large-$N$
expansion used here when applied at the end for systems with as small
$N$.  Extending our analysis such as to compute the next order terms
in the $1/N$-expansion is generally a difficulty task. But we can take
as a basis of the reliability of our results by making a comparison
with other works that have already studied the  phase diagram of the
GN model using alternative approaches to the  large-$N$
expansion. {}For example, in Ref.~\cite{Kneur:2007vm}, the phase
structure of the GN model in (2+1)-dimensions was studied in the
context of the optimized perturbation theory method and which
effectively produces results going beyond the large-$N$ expansion. The
results obtained in that reference show that both the critical
temperature and critical chemical potential are enhanced by a factor
$4N/(4N-1)$, which for $N=2$ means a difference of around $14\%$
percent with respect to the large-$N$ results. Most importantly, those
results also have shown that the phase diagram as a whole is
qualitatively similar to the one obtained in the large-$N$ limit,
except by a small dislocation of the tricritical point to the
$(T,\mu)$ plane, while in the large-$N$ limit it is located at $(T=0,
\mu=\mu_c)$. Even though we do not expect the large-$N$  approximation
to produce precise results for such small values of $N$, like $N=2$ as
in most of the physical systems of interest, it still produces results
of sufficient qualitative agreement and that can provide useful
insights on the physics of these systems. This qualitative agreement
of the large-$N$ approximation is also seen in other works (e.g.,
Refs.~\cite{Kneur:2007vj,Kneur:2013cva}).

\section{The effect of an external magnetic field}
\label{sec5}

Let us now discuss some expected effects of coupling the model we have
studied to the electromagnetic field.  The photon gauge field $A_\mu$
couples to the fermions through the standard quantum electrodynamics
interaction,  $e \bar{\psi} \gamma^\mu \psi A_\mu$. The photon gauge
field, when integrated out, produces the Coulomb interaction between
the fermions.  Note that at the fundamental level, the electromagnetic
gauge field can contribute to our results for the effective potential
through the fermion polarization term coming from the photon gauge
field and fermion interaction. These fermion polarization
contributions are, however, subleading in the large-$N$
approximation. However, the electromagnetic field can (and will be in
general) important when acting as an external source.  In particular,
this can be the case of an external magnetic field applied to the
system.

Let us now briefly discuss some possible consequences of adding an
external magnetic field to the system.  Since the electrons in the 2D
Dirac (Weyl) semimetal are charged, we can add a coupling with the
electromagnetic sector through the minimal coupling, i.e.,
$\partial_\mu \rightarrow \nabla_\mu = \partial_\mu - i e A_\mu$. The
Lagrangian density Eq.~\eqref{lagr1} is then modified as follows,
\begin{eqnarray}\nonumber
\mathcal{L}_0 &=& \bar{\psi} \left( i  M^{\mu \nu}\gamma_\mu
\nabla_\nu \right)  \psi  \\&=& \bar{\psi} \left( i   M^{\mu
  \nu}\gamma_\mu \partial_\nu   \right)  \psi  +e  \bar{\psi}
\gamma_\mu \psi \tilde{A}^\mu,
\label{L0B}
\end{eqnarray}
where we have defined an effective electromagnetic potential
$\tilde{A}_\mu$ with components
\begin{equation}
\tilde{A}_0 =  A_0 -v_F {\bf t} \cdot {\bf A},\;\;\;\; \tilde{A}_i =
\xi_{ij} A_j,\;\;\; i = x,y .
\label{Amu}
\end{equation}
where $\xi_{ij} = \xi_i \delta_{ij}$. Thus, by considering the 2D
system in the $(x,y)$ plane and in the presence of an external and
constant magnetic field ${\bf B}$, we can, without loss of generality,
choose a gauge such that the external four-vector  potential is given
by $A_0=0$ and ${\bf A}_{3D} = - \frac{1}{2} x\hat{\bf x} \times {\bf
  B}$. This leads to a magnetic-field component perpendicular to the
system's plane, $B_{\perp}\equiv B_z$ and another component that is
parallel to the plane, $B_{\parallel}\equiv B_y$.  Hence, from the
tilt vector defined by ${\bf t}=(t_x,t_y,0)$, it is easy to see that
Eq.~(\ref{Amu}) can be expressed in the form
\begin{equation}
\tilde{A}_0 = -\frac{1}{2} v_F x \hat{\bf x} \cdot ({\bf B} \times
      {\bf t} ) = \frac{1}{2} v_F x t_x B_{\perp}. 
\end{equation} 
Therefore, $B_{\perp}$ generates a non-null contribution to the
time component of the effective vector potential $\tilde{A}_0$ and,
thus, the magnetic field will act analogous to an effective external
electric field, with magnitude $\tilde{{\bf E}}_i = \frac{1}{2}v_F
\xi_{ij}({\bf B} \times {\bf t})_j$. Besides this, the anisotropy
in the spatial direction modifies the  magnetic field  in the
perpendicular direction, $\tilde{B}_{\perp} = \xi_x \xi_y B_{\perp}$,
while in the parallel direction, $\tilde{B}_\parallel=  \xi_y
B_\parallel$. Thus, the external magnetic field generates an
electrochemical potential given by 
\begin{equation}
\tilde{\mu}_s = \mu +  \frac{e}{2} v_F x\hat{\bf x}\cdot({\bf B}
\times {\bf t} )  +  \frac{s}{2} g \mu_B |\tilde{{\bf B}}|, 
\end{equation}
where $g$ is the spectroscopic Land\'e factor of the electrons ($g
\approx 2$ in graphene), $s=\pm 1$, and $\mu_B$ is the Bohr magneton. 

We can then see that the tilting of the Dirac cone and the anisotropy
generates two effects. {}First, the combination of an external
magnetic field and the tilt vector ${\bf t}$ generates a non-null
chemical affinity $A_i \propto -{\bf \nabla}_i \mu_s = e ({\bf
  \tilde{E}})_i=   \frac{e}{2}v_F B_{\perp} \xi_{ij} \epsilon^{jk}
t_k$ and, therefore, by Onsager reciprocal relations, it will generate
a current perpendicular to ${\bf t}$. This current  will be
proportional to $\Phi = \int d^2x B_{\perp}$. Hence, one can affirm
that the 2D Dirac (Weyl) semimetal will have an  anomalous Hall
effect~\cite{Burkov:2015hba}. Second, the magnetic-field coupling
generates an anisotropic Zeeman effect and, therefore, it can be
interpreted as an effective giromagnetic term dependent on the
direction of the ${\bf B}_{\parallel}$ field.  It would be worthwhile
to further explore these effects in applications of these type of
planar models when subjected to an external magnetic field and which
we leave as future work.

\section{Conclusions}
\label{conclusions}

In this paper, we have analyzed the problem of chiral symmetry breaking
in a planar four-fermion model when considered in the context of the
generalized Weyl Hamiltonian. The GN-like four-fermion
interaction provides a way to study the chiral symmetry breaking in
the model.  The free Hamiltonian that we have employed in this paper
was the generalized Weyl one, which includes the anisotropies and the
explicit tilting of the Dirac cone. We have then studied how the
tilting of the Dirac cone, parameterized by the so-called effective
tilt parameter $|\tilde{\bf t}|$, affects the chiral phase transition
as a function of both chemical potential (e.g., doping) and finite
temperature. Both the critical temperature and critical chemical
potential decrease by a factor $\sqrt{1-|\tilde{\bf t}|^2}$ with
respect to the untilted case.  Thus, the tilt parameter suppresses the
chiral symmetry breaking, easing its restoration.

{}Finally, we have shown that the planar fermion system that we have
studied will respond to an external magnetic field differently when in
the absence or in the presence of the Dirac cone tilt. In the
anisotropic  and tilted case, there will be an anomalous Hall effect
and its magnitude is proportional to the Lorentz-violating parameters
${\bm \xi}$ and ${\bf t}$. This effect changes the effective chemical
potential and a current perpendicular to  ${\bf t}$ appears. The
anomalous Hall effect~\cite{Burkov:2015hba} is related to the
emergence of this unusual current ${\bf j} \propto {\bf t} \times {\bf
  B}$. In the model case we have studied, this effect comes from the
explicit  Lorentz-violating structure of the matrix $M^{\mu \nu}$ in
Eq.~(\ref{L0B}).  It generates a non-constant contribution to the
chemical potential, which is an explicit character of
non-equilibrium systems and which can lead to interesting features in
the study of transport phenomena.  These and other features found here
as a consequence of this paper will be analyzed in more details in a
future work.

\section*{Acknowledgments}

Y.M.P.G. is supported by a postdoctoral grant from  {}Funda\c{c}\~ao
Carlos Chagas Filho de Amparo \`a Pesquisa do Estado do Rio de Janeiro
(FAPERJ).  R.O.R. is partially supported by research grants from
Conselho Nacional de Desenvolvimento Cient\'{\i}fico e Tecnol\'ogico
(CNPq), Grant No. 302545/2017-4, and from {}Funda\c{c}\~ao
Carlos Chagas Filho de Amparo \`a Pesquisa do Estado do Rio de Janeiro
(FAPERJ), Grant No. E-26/201.150/2021.
 


\end{document}